\documentclass[12pt]{article}

\usepackage{amssymb}

\usepackage[dvips]{graphicx}

\begin{document}

\title
{Calculation of two-loop $\beta$-function for general N=1
supersymmetric Yang--Mills theory with the higher covariant
derivative regularization}

\author{A.B.Pimenov, E.S.Shevtsova, K.V.Stepanyantz}

\maketitle

\begin{center}

{\em Moscow State University, physical faculty,\\
department of theoretical physics.\\
$119992$, Moscow, Russia}

\end{center}

\begin{abstract}
For the general renormalizable N=1 supersymmetric Yang--Mills
theory, regularized by higher covariant derivatives, a two-loop
$\beta$-function is calculated. It is shown that all integrals,
needed for obtaining this function, can be easily calculated,
because they are integrals of total derivatives.
\end{abstract}


\section{Introduction.}
\hspace{\parindent}

It is well known that most quantum field theory models are
divergent in the ultraviolet region. In order to deal with the
divergent expressions, it is necessary to regularize a theory.
Although physical results does not depend on regularization, a
proper choice of the regularization can considerably simplify
calculations or reveal some features of quantum corrections. Most
calculations in the quantum field theory where made with the
dimensional regularization \cite{tHooft}. However, the dimensional
regularization is not convenient for calculations in
supersymmetric theories, because it breaks the supersymmetry. That
is why in supersymmetric theories one usually uses its
modification, called the dimensional reduction \cite{Siegel}.
There are a lot of calculation, made in supersymmetric theories
with the dimensional reduction, see e.f. \cite{DRED}. However, it
is well known that the dimensional reduction is not
self-consistent \cite{Siegel2}. Ways, allowing to avoid such
problems, are discussed in the literature \cite{Stockinger}. Other
regularizations are sometimes applied for calculations in
supersymmetric theories. For example, in Ref. \cite{Mas} two-loop
$\beta$-function of the N=1 supersymmetric Yang--Mills theory was
calculated with the differential renormalization \cite{DiffR}.

A self-consistent regularization, which does not break the
supersymmetry, is the higher covariant derivative regularization
\cite{Slavnov}, which was generalized to the supersymmetric case
in Ref. \cite{Krivoshchekov} (another variant was proposed in Ref.
\cite{West_Paper}). However, using this regularization is rather
technically complicated. The first calculation of quantum
corrections for the (non-supersymmetric) Yang--Mills theory was
made in Ref. \cite{Ruiz}. Taking into account corrections, made in
subsequent papers \cite{Asorey}, the result for the
$\beta$-function appeared to be the same as the well-known result,
obtained with the dimensional regularization \cite{Politzer}. In
principle, it is possible to prove that in the one-loop
approximation calculations with the higher covariant derivative
regularization always agree with the results of calculations with
the dimensional regularization \cite{PhysLett}. Some calculations
in the one-loop and two-loop approximations were made for various
theories \cite{Rosten_Calc,Rosten_Int} with a variant of the
higher covariant derivative regularization, proposed in
\cite{Rosten_Reg}. The structure of the corresponding integrals
was discussed in Ref. \cite{Rosten_Int}.

Application of the higher covariant derivative regularization to
calculation of quantum corrections in the N=1 supersymmetric
electrodynamics in two and three loops \cite{3LoopHEP,Review}
reveals an interesting feature of quantum corrections: all
integrals, defining the $\beta$-function appear to be integrals of
total derivatives and can be easily calculated. This makes
possible analytical multiloop calculations with the higher
covariant derivative regularization in supersymmetric theories and
allows to explain the origin of the NSVZ $\beta$-function, which
relates the $\beta$-function in $n$-th loop with the
$\beta$-function and the anomalous dimensions in the previous
loops. Due to this, application of this regularization is
sometimes very convenient in the supersymmetric case. The fact
that the integrals, appearing with the higher covariant derivative
regularization, in the limit of zero external momentum become
integrals of total derivatives, seems to be a general feature of
all supersymmetric theories. Nevertheless, with the higher
derivative regularization even the two-loop $\beta$-function has
not yet been calculated for a general N=1 supersymmetric
Yang--Mills theory. This is made in this paper. Note that in order
to do this calculation, it is necessary to introduce higher
covariant derivative terms not only for the gauge field, but also
for the matter superfields.

The paper is organized as follows:

In Sec. \ref{Section_SUSY} we introduce the notation and recall
basic information about the higher covariant derivative
regularization. The $\beta$-function for the considered theory is
calculated in Sec. \ref{Section_Matter_Contribution}. The result
is briefly discussed in the Conclusion.


\section{N=1 supersymmetric Yang--Mills theory and the higher covariant
derivative regularization}
\hspace{\parindent}\label{Section_SUSY}

In this paper we calculate $\beta$-function for a general
renormalizable N=1 supersymmetric Yang--Mills theory. In the
massless case this theory is described by the action

\begin{eqnarray}\label{SYM_Action}
&& S = \frac{1}{2 e^2} \mbox{Re}\,\mbox{tr}\int
d^4x\,d^2\theta\,W_a C^{ab} W_b + \frac{1}{4}\int d^4x\,
d^4\theta\, (\phi^*)^i (e^{2V})_i{}^j\phi_j +\nonumber\\
&&\qquad\qquad\qquad\qquad\qquad\qquad\qquad\qquad +
\Bigg(\frac{1}{6} \int d^4x\, d^2\theta\,\lambda^{ijk} \phi_i
\phi_j \phi_k + \mbox{h.c.}\Bigg),\qquad
\end{eqnarray}

\noindent where $\phi_i$ are chiral matter superfields in a
representation $R$, which is in general reducible. $V$ is a real
scalar gauge superfield. The superfield $W_a$ is a supersymmetric
gauge field stress tensor, which is defined by

\begin{equation}
W_a = \frac{1}{8} \bar D^2 (e^{-2V} D_a e^{2V}).
\end{equation}

\noindent In our notation $D_a$ and $\bar D_a$ are the right and
left supersymmetric covariant derivatives respectively, $V = e\,
V^A T^A$, and the generators of the fundamental representation are
normalized by the condition

\begin{equation}
\mbox{tr}(t^A t^B) = \frac{1}{2}\delta^{AB}.
\end{equation}

\noindent Action (\ref{SYM_Action}) should be invariant under the
gauge transformations

\begin{equation}
\phi \to e^{i\Lambda}\phi;\qquad e^{2V} \to e^{i\Lambda^+} e^{2V}
e^{-i\Lambda},
\end{equation}

\noindent where $\Lambda$ is an arbitrary chiral superfield. As a
consequence, the coefficient $\lambda^{ijk}$ should satisfy the
condition

\begin{equation}\label{Lambda_Invariance}
(T^A)_m{}^{i}\lambda^{mjk} + (T^A)_m{}^{j}\lambda^{imk} +
(T^A)_m{}^{k}\lambda^{ijm} = 0.
\end{equation}

For calculation of quantum corrections it is convenient to use the
background field method. In the supersymmetric case it can be
formulated as follows \cite{West}: Let us make the substitution

\begin{equation}\label{Substitution}
e^{2V} \to e^{2V'} \equiv e^{\mbox{\boldmath${\scriptstyle
\Omega}$}^+} e^{2V} e^{\mbox{\boldmath${\scriptstyle \Omega}$}},
\end{equation}

\noindent in action (\ref{SYM_Action}), where
$\mbox{\boldmath${\Omega}$}$ is a background superfield. Then the
theory is invariant under the background gauge transformations

\begin{equation}\label{Background_Transformations}
\phi \to e^{i\Lambda}\phi;\quad  V \to e^{iK} V e^{-iK}; \quad
e^{\mbox{\boldmath${\scriptstyle \Omega}$}} \to e^{iK}
e^{\mbox{\boldmath${\scriptstyle \Omega}$}} e^{-i\Lambda};\quad
e^{\mbox{\boldmath${\scriptstyle \Omega}$}^+} \to e^{i\Lambda^+}
e^{\mbox{\boldmath${\scriptstyle \Omega}$}^+} e^{-iK},
\end{equation}

\noindent where $K$ is an arbitrary real superfield, and $\Lambda$
is a background-chiral superfield. This invariance allows to set
$\mbox{\boldmath$\Omega$} = \mbox{\boldmath$\Omega$}^+ = {\bf V}$.

It is convenient to choose a regularization and gauge fixing so
that invariance (\ref{Background_Transformations}) is unbroken.
First, we fix a gauge by adding

\begin{equation}\label{Gauge_Fixing}
S_{\mbox{\scriptsize gf}} = - \frac{1}{32 e^2}\,\mbox{tr}\,\int
d^4x\,d^4\theta\, \Big(V \mbox{\boldmath$D$}^2
\bar{\mbox{\boldmath$D$}}^2  V + V \bar {\mbox{\boldmath$D$}}^2
\mbox{\boldmath$D$}^2 V\Big)
\end{equation}

\noindent to the action. The corresponding Faddeev--Popov and
Nielsen--Kallosh ghost Lagrangians are constructed by the standard
way.

For regularization we add the terms

\begin{eqnarray}\label{Regularized_Action}
&& S_\Lambda = \frac{1}{2 e^2}\mbox{tr}\,\mbox{Re}\int
d^4x\,d^4\theta\, V\frac{(\mbox{\boldmath$D$}_\mu^2)^{n+1}}{
\Lambda^{2n}} V + \frac{1}{8} \int
d^4x\,d^4\theta\,\Bigg((\phi^*)^i
\Big[e^{\mbox{\boldmath${\scriptstyle \Omega}$}^+} e^{2V}
\frac{(\mbox{\boldmath$D$}_\alpha^2)^{m}}{\Lambda^{2m}}
e^{\mbox{\boldmath${\scriptstyle \Omega}$}}\Big]{}_i{}^j \phi_j
+\nonumber\\
&& + (\phi^*)^i \Big[e^{\mbox{\boldmath${\scriptstyle \Omega}$}^+}
\frac{(\mbox{\boldmath$D$}_\alpha^2)^{m}}{\Lambda^{2m}} e^{2V}
e^{\mbox{\boldmath${\scriptstyle \Omega}$}}\Big]{}_i{}^j
\phi_j\Bigg),
\end{eqnarray}

\noindent where $\mbox{\boldmath$D$}_\alpha$ is the background
covariant derivative and we assume that $m<n$.\footnote{Other
choices of the higher derivative terms are also possible.}
(Because the considered theory contains a nontrivial
superpotential, it is also necessary to introduce the higher
covariant derivative term for the matter superfields.)

The regularized theory is evidently invariant under the background
gauge transformations. The regularization, described above, is
rather simple, but breaks the BRST-invariance of the action. That
is why it is necessary to use a special subtraction scheme, which
restore the Slavnov--Taylor identities in each order of the
perturbation theory \cite{Slavnov12}. For the supersymmetric case
such a scheme was constructed in Ref. \cite{Slavnov34}.

It is well-known \cite{Slavnov_Book} that the higher covariant
derivative term does not remove divergences in the one-loop
approximation. In order to cancel the remaining one-loop
divergences, it is necessary to introduce into the generating
functional the Pauli--Villars determinants

\begin{equation}
\prod\limits_{I}\Big(\int D\phi_{I}^* D\phi_{I}
e^{iS_{I}}\Big)^{-c_I},
\end{equation}

\noindent where $S_I$ is the action for the Pauli--Villars
fields,\footnote{Note that this action differs from the one, used
in \cite{3LoopHEP}, because here the quotient of the coefficients
in the kinetic term and in the mass term does not contain the
factor $Z$. Using terminology of Ref. \cite{Arkani}, one can say
that here we calculate the canonical coupling $\alpha_c$, while in
Ref. \cite{3LoopHEP} we calculated the holomorphic coupling
$\alpha_h$. Certainly, after the renormalization the effective
action does not depend on the definitions. However, the
definitions used here are much more convenient.}

\begin{eqnarray}
&& S_I = \frac{1}{8} \int d^4x\,d^4\theta\,\Bigg((\phi_I^*)^i
\Big[e^{\mbox{\boldmath${\scriptstyle \Omega}$}^+} e^{2V}
\Big(1+\frac{(\mbox{\boldmath$D$}_\alpha^2)^{m}}{\Lambda^{2m}}\Big)
e^{\mbox{\boldmath${\scriptstyle \Omega}$}}\Big]{}_i{}^j
(\phi_I)_j + (\phi_I^*)^i \Big[e^{\mbox{\boldmath${\scriptstyle
\Omega}$}^+}
\Big(1+\frac{(\mbox{\boldmath$D$}_\alpha^2)^{m}}{\Lambda^{2m}}
\Big)\times\nonumber\\
&& \times e^{2V} e^{\mbox{\boldmath${\scriptstyle
\Omega}$}}\Big]{}_i{}^j (\phi_I)_j\Bigg) + \Big(\frac{1}{4}\int
d^4x\,d^2\theta\,M_I^{ij} (\phi_{I})_i (\phi_{I})_j +\mbox{h.c.}
\Big).
\end{eqnarray}

\noindent The masses of the Pauli--Villars fields are proportional
to the parameter $\Lambda$:

\begin{equation}
M^{ij}_I = a_I^{ij}\Lambda.
\end{equation}

\noindent This means that $\Lambda$ is the only dimensionful
parameter of the regularized theory. We assume that the mass term
does not break the gauge invariance. Also we will choose the
masses so that

\begin{equation}
M_I^{ij} (M_I^*)_{jk} = M_I^2 \delta_k^i.
\end{equation}

The coefficients $c_I$ satisfy the conditions

\begin{equation}
\sum\limits_I c_I = 1;\qquad \sum\limits_I c_I M_I^2 = 0.
\end{equation}

The generating functional for connected Green functions and the
effective action are defined by the standard way.

In this paper we will calculate the $\beta$-function. We use the
following notation. Terms in the effective action, corresponding
to the renormalized two-point Green function of the gauge
superfield, are written as

\begin{equation}\label{D_Definition}
\Gamma^{(2)}_V = - \frac{1}{8\pi} \mbox{tr}\int
\frac{d^4p}{(2\pi)^4}\,d^4\theta\,{\bf V}(-p)\,\partial^2\Pi_{1/2}
{\bf V}(p)\, d^{-1}(\alpha,\lambda,\mu/p).
\end{equation}

\noindent where $\alpha$ is a renormalized coupling constant. We
calculate

\begin{equation}
\frac{d}{d\ln \Lambda}\,
\Big(d^{-1}(\alpha_0,\lambda_0,\Lambda/p)-\alpha_0^{-1}\Big)\Big|_{p=0}
= - \frac{d\alpha_0^{-1}}{d\ln\Lambda} =
\frac{\beta(\alpha_0)}{\alpha_0^2}.
\end{equation}

\noindent The anomalous dimension is defined similarly. First we
consider the two-point Green function for the matter superfield in
the massless limit:

\begin{equation}\label{Renormalized_Gamma_2}
\Gamma^{(2)}_\phi = \frac{1}{4} \int
\frac{d^4p}{(2\pi)^4}\,d^4\theta\,(\phi^*)^i(-p,\theta)\,\phi_j(p,\theta)
\,(ZG)_i{}^j(\alpha,\lambda,\mu/p),
\end{equation}

\noindent where $Z$ denotes the renormalization constant for the
matter superfield. Then the anomalous dimensions is defined by

\begin{equation}
\gamma_i{}^j\Big(\alpha_0(\alpha,\lambda,\Lambda/\mu)\Big) =
-\frac{\partial}{\partial\ln \Lambda} \Big(\ln
Z(\alpha,\lambda,\Lambda/\mu)\Big)_i{}^j.
\end{equation}


\section{Two-loop $\beta$-function}
\label{Section_Matter_Contribution} \hspace{\parindent}

After calculation of the supergraphs, we have obtained the
following result for the two-loop $\beta$-function:

\begin{eqnarray}\label{Result}
&& \beta_2(\alpha) = -\frac{3\alpha^2}{2\pi} C_2 + \alpha^2 T(R)
I_0 + \alpha^3 C_2^2 I_1 + \frac{\alpha^3}{r} C(R)_i{}^j
C(R)_j{}^i I_2 + \alpha^3 T(R) C_2 I_3 +\quad\nonumber\\
&& + \alpha^2 C(R)_i{}^j \frac{\lambda_{jkl}^* \lambda^{ikl}}{4\pi
r} I_4,
\end{eqnarray}

\noindent where the following notation is used:

\begin{eqnarray}\label{T(R)}
&& \mbox{tr}\,(T^A T^B) \equiv T(R)\,\delta^{AB};\qquad
(T^A)_i{}^k
(T^A)_k{}^j \equiv C(R)_i{}^j;\nonumber\\
&& f^{ACD} f^{BCD} \equiv C_2 \delta^{AB};\qquad\quad r\equiv
\delta_{AA}.
\end{eqnarray}

\noindent (Note that $T(R) = C(R)_i{}^i/r$.) Here

\begin{equation}
I_i = I_i(0) -\sum\limits_{I} c_I I_i(M_I)\quad\mbox{for}\quad
I=0,2,3,
\end{equation}

\noindent and the integrals $I_0(M)$, $I_1$, $I_2(M)$, $I_3(M)$
and $I_4$ are given by

\begin{eqnarray}
&& I_0(M) = 4\pi \int \frac{d^4q}{(2\pi)^4} \frac{d}{d\ln\Lambda}
\frac{1}{q^2} \frac{d}{dq^2}
\Bigg[\ln\Big(q^2(1+q^{2m}/\Lambda^{2m})^2+M^2\Big)
+\nonumber\\
&& + \frac{M^2}{q^2(1+q^{2m}/\Lambda^{2m})^2 +M^2} - \frac{2m\,
q^{2m}/\Lambda^{2m}
q^2(1+q^{2m}/\Lambda^{2m})}{q^2(1+q^{2m}/\Lambda^{2m})^2+M^2}
\Bigg];\\
&&\vphantom{1}\nonumber\\
&& I_1 = 96\pi^2 \int \frac{d^4q}{(2\pi)^4} \frac{d^4k}{(2\pi)^4}
\frac{d}{d\ln\Lambda} \frac{1}{k^2}
\frac{d}{dk^2}\Bigg[\frac{1}{q^2 (q+k)^2 (1+q^{2n}/\Lambda^{2n})
(1+(q+k)^{2n}/\Lambda^{2n})}\times\nonumber\\
&& \times \Bigg(\frac{n+1}{(1+k^{2n}/\Lambda^{2n})} - \frac{n}{
(1+k^{2n}/\Lambda^{2n})^{2}}\Bigg) \Bigg];\\
&&\vphantom{1}\nonumber\\
&& I_2(M) = - 16\pi^2 \int \frac{d^4q}{(2\pi)^4}
\frac{d^4k}{(2\pi)^4} \frac{d}{d\ln\Lambda} \frac{1}{q^2}
\frac{d}{dq^2} \frac{(1+ (q+k)^{2m}/\Lambda^{2m})}{\Big((q+k)^2(1+
(q+k)^{2m}/\Lambda^{2m})^2 +M^2\Big)}\times\vphantom{\Bigg(}\nonumber\\
&& \times \frac{1}{k^2 (1+k^{2n}/\Lambda^{2n})}\Bigg[ \frac{q^4
(2+(q+k)^{2m}/\Lambda^{2m}+ q^{2m}/\Lambda^{2m})^2
(1+q^{2m}/\Lambda^{2m})^3}{\Big(q^2(1+q^{2m}/\Lambda^{2m})^2
+M^2\Big)^2}
+\vphantom{\Bigg(}\nonumber\\
&& + m q^{2m}/\Lambda^{2m} \Bigg(- \frac{2q^2
(2+(q+k)^{2m}/\Lambda^{2m}+
q^{2m}/\Lambda^{2m})(1+q^{2m}/\Lambda^{2m})}{q^2(1+q^{2m}/\Lambda^{2m})^2
+M^2} +\vphantom{\Bigg(}\\
&& + \frac{q^2 (2+(q+k)^{2m}/\Lambda^{2m}+
q^{2m}/\Lambda^{2m})^2}{q^2(1+q^{2m}/\Lambda^{2m})^2 +M^2} -
\frac{2q^2 M^2 (2+(q+k)^{2m}/\Lambda^{2m}+
q^{2m}/\Lambda^{2m})^2}{\Big(q^2(1+q^{2m}/\Lambda^{2m})^2
+M^2\Big)^2}
\Bigg) \Bigg];\nonumber\\
&&\vphantom{1}\nonumber\\
&& I_3(M) = 4\pi^2 \int \frac{d^4q}{(2\pi)^4}
\frac{d^4k}{(2\pi)^4} \frac{d}{d\ln\Lambda}\Bigg\{
\frac{\partial}{\partial q_\alpha} \Bigg[\frac{k_\alpha}{(k+q)^2
(1+ (q+k)^{2n}/\Lambda^{2n})}
\times \vphantom{\Bigg(}\nonumber\\
&& \times \Bigg(- \frac{(2+ k^{2m}/\Lambda^{2m} +
q^{2m}/\Lambda^{2m})^2(1+k^{2m}/\Lambda^{2m})^3
(1+q^{2m}/\Lambda^{2m})}{\Big(k^2(1+
k^{2m}/\Lambda^{2m})^2+M^2\Big)^2
\Big(q^2(1+ q^{2m}/\Lambda^{2m})^2+M^2\Big)} -\nonumber\\
&& - \frac{m\, k^{2m}/\Lambda^{2m}(2+ k^{2m}/\Lambda^{2m} +
q^{2m}/\Lambda^{2m})^2 (1+q^{2m}/\Lambda^{2m})}{k^2 \Big(k^2(1+
k^{2m}/\Lambda^{2m})^2+M^2\Big) \Big(q^2(1+
q^{2m}/\Lambda^{2m})^2+M^2\Big)}
+\vphantom{\Bigg(}\nonumber\\
&& + \frac{2m\, k^{2m}/\Lambda^{2m}(2+ k^{2m}/\Lambda^{2m} +
q^{2m}/\Lambda^{2m})
(1+k^{2m}/\Lambda^{2m})(1+q^{2m}/\Lambda^{2m})}{k^2 \Big(k^2(1+
k^{2m}/\Lambda^{2m})^2+M^2\Big) \Big(q^2(1+
q^{2m}/\Lambda^{2m})^2+M^2\Big)}
+\vphantom{\Bigg(}\nonumber\\
&& + \frac{2m\, M^2 k^{2m}/\Lambda^{2m}(2+ k^{2m}/\Lambda^{2m} +
q^{2m}/\Lambda^{2m})^2 (1+q^{2m}/\Lambda^{2m})}{k^2 \Big(k^2(1+
k^{2m}/\Lambda^{2m})^2+M^2\Big)^2 \Big(q^2(1+
q^{2m}/\Lambda^{2m})^2+M^2\Big)}
\Bigg)\Bigg] -\nonumber\\
&& - \frac{1}{k^2} \frac{d}{dk^2} \Bigg[\frac{2(2+
(q+k)^{2m}/\Lambda^{2m} + q^{2m}/\Lambda^{2m})^2
(1+q^{2m}/\Lambda^{2m})(1+(q+k)^{2m}/\Lambda^{2m})}{
\Big(q^2(1+q^{2m}/\Lambda^{2m})^2+M^2\Big)
\Big((q+k)^2(1+(q+k)^{2m}/\Lambda^{2m})^2 +
M^2\Big)}\times\nonumber\\
&& \times \Bigg(\frac{1}{(1+k^{2n}/\Lambda^{2n})} + \frac{n
k^{2n}/\Lambda^{2n}}{(1+k^{2n}/\Lambda^{2n})^2}\Bigg) \Bigg]\Bigg\};\nonumber\\
&&\vphantom{1}\nonumber\\
&& I_4 = 64\pi^2 \int \frac{d^4q}{(2\pi)^4} \frac{d^4k}{(2\pi)^4}
\frac{d}{d\ln\Lambda} \frac{1}{q^2} \frac{d}{dq^2}
\Bigg[\frac{1}{k^2 (q+k)^2
(1+k^{2m}/\Lambda^{2m})}\times\nonumber\\
&& \times \frac{1}{(1+(q+k)^{2m}/\Lambda^{2m})}
\Bigg(\frac{1}{(1+q^{2m}/\Lambda^{2m})} + \frac{m
q^{2m}/\Lambda^{2m}}{(1+q^{2m}/\Lambda^{2m})^2}\Bigg) \Bigg].
\end{eqnarray}

It is easy to see that all these integrals are integrals of total
derivatives, due to the identity

\begin{equation}\label{Integral_Of_Total_Derivative}
\int \frac{d^4q}{(2\pi)^4}\frac{1}{q^2} \frac{d}{dq^2} f(q^2) =
\frac{1}{16\pi^2} \Big(f(q^2=\infty) - f(q^2=0)\Big),
\end{equation}

\noindent which can be easily proved in the four-dimensional
spherical coordinates. Using this identity we find

\begin{eqnarray}
&& I_0 = \frac{1}{4\pi} \frac{d}{d\ln\Lambda}\Big(\sum\limits_I
c_I \ln M_I^2 \Big) = \frac{1}{2\pi};\nonumber\\
&& I_1 = - 6 \int \frac{d^4q}{(2\pi)^4} \frac{d}{d\ln\Lambda}
\Bigg[\frac{1}{q^4 (1+q^{2n}/\Lambda^{2n})^2} \Bigg] =
- \frac{3}{4\pi^2};\nonumber\\
&& I_2 = \int \frac{d^4k}{(2\pi)^4} \frac{d}{d\ln\Lambda}
\Bigg[\frac{(2+k^{2m}/\Lambda^{2m})^2}{k^4
(1+k^{2n}/\Lambda^{2n})(1+k^{2m}/\Lambda^{2m})} \Bigg] =
\frac{1}{2\pi^2};\nonumber\\
&& I_3 = \int \frac{d^4q}{(2\pi)^4} \frac{d}{d\ln\Lambda}
\Bigg[\frac{2}{q^4} -\sum\limits_I c_I
\frac{2(1+q^{2m}/\Lambda^{2m})^4}{(q^2(1+q^{2m}/\Lambda^{2m})^2
+ M_I^2)^2} \Bigg] = \frac{1}{4\pi^2};\nonumber\\
&& I_4 = - \int \frac{d^4k}{(2\pi)^4} \frac{d}{d\ln\Lambda}
\Bigg[\frac{4}{k^4 (1+k^{2m}/\Lambda^{2m})^2} \Bigg] =
-\frac{1}{2\pi^2}.
\end{eqnarray}

\noindent Note that the Pauli--Villars fields nontrivially
contributes only to integrals $I_0$ and $I_3$, where they are very
important. For example, in the two-loop integral $I_3$ the
Pauli--Villars contribution cancels the one-loop subdivergence,
produced by the matter superfields.

Thus, in the two-loop approximation

\begin{eqnarray}
&& \beta(\alpha) = - \frac{\alpha^2}{2\pi}\Big(3 C_2 - T(R)\Big) +
\frac{\alpha^3}{(2\pi)^2}\Big(-3 C_2^2 + T(R) C_2 + \frac{2}{r}
C(R)_i{}^j C(R)_j{}^i\Big) -\nonumber\\
&& - \frac{\alpha^2 C(R)_i{}^j \lambda_{jkl}^*
\lambda^{ikl}}{8\pi^3 r} + \ldots
\end{eqnarray}

\noindent Taking into account that the one-loop anomalous
dimension is given by

\begin{equation}
\gamma_i{}^j(\alpha) = -\frac{\alpha C(R)_i{}^j}{\pi} +
\frac{\lambda_{ikl}^* \lambda^{jkl}}{4\pi^2} + \ldots,
\end{equation}

\noindent we see that our result agrees with the exact NSVZ
$\beta$-function \cite{NSVZ_Instanton}

\begin{equation}\label{NSVZ_Beta}
\beta(\alpha) = - \frac{\alpha^2\Big[3 C_2 - T(R) + C(R)_i{}^j
\gamma_j{}^i(\alpha)/r\Big)\Big]}{2\pi(1- C_2\alpha/2\pi)}.
\end{equation}

\noindent Up to notation, this result is in agreement with the
results of calculations made with the dimensional reduction, see
e.f. \cite{DRED}.


\section{Conclusion}
\label{Section_Conclusion} \hspace{\parindent}

In this paper we demonstrate, how the two--loop $\beta$-function
in N=1 supersymmetric theories can be calculated with the higher
covariant derivative regularization. The most interesting feature
of this calculation is the factorization of rather complicated
integrals into integrals of total derivatives. Partially this fact
can be explained substituting solutions of Slavnov--Taylor
identities into the Schwinger--Dyson equations. However, a
complete proof of this fact has not yet been done. Its origin is
also so far unclear. Possibly, this feature appears due to using
of the background field method \cite{Smilga}. Factorization of
integrals, obtained with the higher covariant derivative
regularization, into integrals of total derivatives can allow to
do a simple derivation of the Novikov, Shifman, Vainshtein, and
Zakharov $\beta$-function, which relates $n$-loop contribution to
the $\beta$-function with the $\beta$-function and the anomalous
dimension in previous loops. In this paper we have shown how this
can be done at the two-loop level.

\bigskip
\bigskip

\noindent {\Large\bf Acknowledgements.}

\bigskip

\noindent This work was partially supported by RFBR grant No
08-01-00281a. K.V.Stepanyantz is very grateful to Dr. O.J.Rosten
for a valuable discussion.


\end{document}